\documentclass[]{article}
\usepackage[left=0.75in,top=1.0in,bottom=1.0in,right=0.75in,nohead]{geometry}

\author{Stuart B. Heinrich \\ sbheinric@gmail.com}
\title{Quantum Fluid Dynamics on the Hypersphere}

\usepackage[round,numbers]{natbib}
\usepackage[pdftex]{xcolor,graphicx}
\usepackage[linkbordercolor=white]{hyperref}
\usepackage{amssymb}
\usepackage[cmex10]{amsmath}
\usepackage{amssymb}
\usepackage{theorem}
\usepackage{verbatim}
\usepackage{url}
\usepackage{cancel}
\usepackage{rotating}
\usepackage{multicol}
\usepackage{algorithm}
\usepackage{algorithmic}
\usepackage{boxedminipage}
\usepackage[labelsep=period,font={footnotesize}]{caption} 
\usepackage{subfig}
\usepackage{cancel}
\usepackage{wrapfig}
\usepackage{textcomp} 
\usepackage[version=4]{mhchem}
\usepackage{seqsplit}

\newcommand{\eqnref}[1]{ (\ref{#1})}

\newcommand{\figref}[1]{Fig. \ref{#1}}
\newcommand{\secref}[1]{Section \ref{#1}}

\newcommand{\pd}[2]{ \frac{\partial #1}{\partial #2} }
\newcommand{\pdd}[2]{ \frac{\partial^2 #1}{\partial #2} }

\DeclareMathOperator{\atan}{atan}

\newcommand{\norm}[1]{ \left| \left| #1 \right| \right| }

\begin{document}
\maketitle

\begin{abstract}
It is known from quantum mechanics that particles are associated with wave functions, and that the probability of observing a particle at some future location is proportional to the squared modulus of the amplitude of its wave function.  Although this statistical relationship is well quantified, the interpretations have remained controversial, with many split between the classical Copenhagen interpretation and some variation of the de Broglie-Bohm pilot wave models.  Recent experiments with Hydrodynamic Quantum Analogs (HQAs) have demonstrated that droplets of real fluid may achieve stable dynamical states, where interaction of the droplets with their own ripples results in motion analogous to the motion of particles subject to the guiding equation under the pilot wave models.  Indeed, many effects previously thought to be exclusively quantum have now been observed as emergent phenomena in these macroscopic HQAs.  This has motivated us to explore the possibility that quantum mechanics may actually be the result of fluid dynamics on some real quantum scale fluid analogous to the fluid in macroscopic HQAs (i.e., where elementary particles are actually bouncing droplets of fluid having properties defined by their dynamics, and the wave functions are ripples produced by the interaction of those bouncing droplets with the fluid surface).  In this paper, we show that if there is a real quantum scale fluid having dynamics analogous to the fluid in HQAs, then this fluid must be a superfluid, and the dynamics of that fluid must take place on the surface of a 4-dimensional hypersphere.  Under the influence of cosmological inflation, we further show that these bouncing droplets would have the illusion of a property analogous to rest mass, and that the principles of inertia, momentum, mass-energy equivalence, general relativity, the uncertainty principle and the appearance of a time-like dimension can all be derived for droplets as purely emergent phenomena from the fluid dynamics of this system.  As such, we believe that this model merits consideration as a potential foundation for a new unifying theory of physics at all scales.
\end{abstract}

\ifx \kindle \undefined
\begin{multicols}{2}
\else
\thispagestyle{empty}
\pagestyle{empty}
\fi

\section{Introduction}

Inspired by early pilot wave models of quantum mechanics originally pioneered by de Broglie \citep{deBroglie1927} and Bohm \citep{Bohm1952,Bohm1952b}, coupled with observations from more recent Hydrodynamic Quantum Analogs (HQAs) \citep{Bush_2020}, we introduce a new theory of Quantum Fluid Dynamics (QFD).  

Although QFD is not a fully developed theory, and the potential for reconciliation with electric charge, spin, and other properties has not yet been investigated, QFD demonstrates remarkable potential to unify certain quantum scale observations with cosmological scale observations from a new perspective, suggesting that quantum mechanics may be intrinsically related to gravitation, inflation and the shape of the universe.

We begin by providing relevant historical context of theories leading up to the development of Hydrodynamic Quantum Field Theory (HQFT) (\secref{sec_background}) before introducing our theory of QFD, which resolves many of the open questions remaining from HQFT (\secref{sec_QFD}).  

We show that mass-energy equivalence can be derived from QFD (\secref{sec_mass_energy_equivalence}), and that the mechanism by which particles have mass, and the theory of gravitation, can be derived from QFD under the influence of cosmological inflation (\secref{sec_gravitation}).  We then show that inertia and momentum can also be derived from QFD (\secref{sec_inertia_and_momentum}), as well as the uncertainty principle (\secref{sec_fluctuations}), the existence of massless particles (\secref{sec_massless}), and the illusion of a time-like dimension (\secref{sec_inflation_and_time}).  Finally, we discuss techniques for numerical simulation  (\secref{sec_simulation}).

\section{Background} \label{sec_background}

In quantum mechanics, it is known that particles are associated with wave functions.  For example, the time evolution of non-relativistic spin-0 particles is governed by Schr\"{o}dinger's equation, and the squared modulus of the amplitude of the wave function corresponds to the probability density of observing the particle at any given future position (known as the Born identity) \citep{griffiths:quantum}.

This relationship between waves and particles is exemplified by the double-slit experiment.  When a single particle such as a photon or an electron passes through a double-slit, its probability of being detected at any position on the detector plate is governed by the interference pattern produced by the associated wave equation after passing through the two slits (\figref{fig_double_slit}).

\begin{figure}[H]
\centering
\includegraphics[width=0.5 \textwidth]{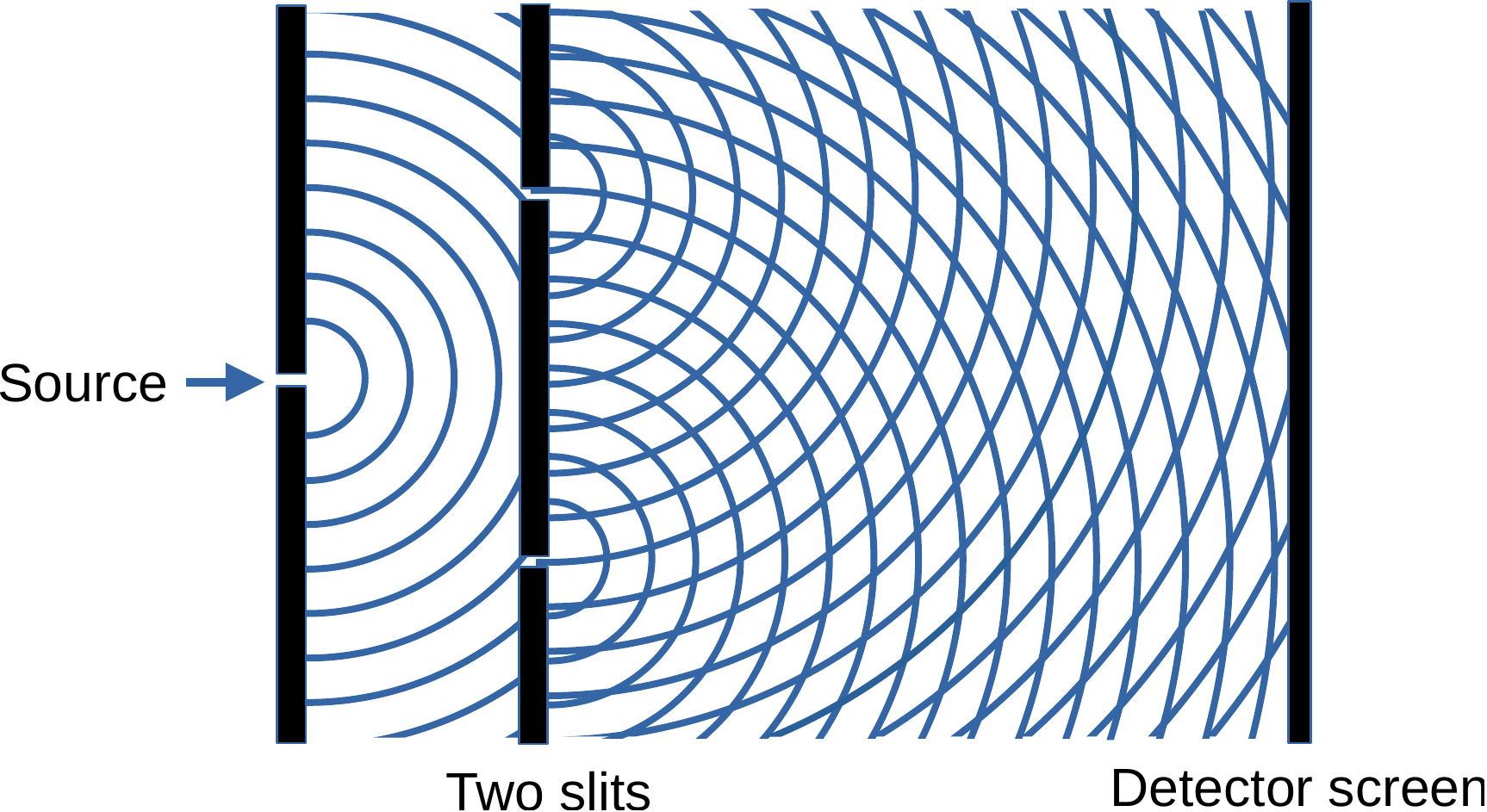}
\caption{Single particles emitted from the source are detected at the screen with probability density according to the interference pattern that would be produced by circular waves.}
  \label{fig_double_slit}
\end{figure}

While the statistical validity of this above description is uncontested, the interpretations remain controversial.  Indeed, there exist multiple competing theories that provide different explanations for this same statistical observation.

Under the Copenhagen interpretation \citep{Omnes1999}, which is the standard interpretation of quantum mechanics, it is believed that this statistical relationship is the ultimate description of reality with no deeper explanation (i.e., no additional ``hidden variables''). In other words, while not being observed, particles are believed to have no definite position, instead existing in a ``superposition'' of all possibilities simultaneously (with probability described by the wave function).  Only when particles are ``observed'' do they suddenly assume a definite position, sampled at random from the probability density described by the wave function.

Under this interpretation, the wave function is not regarded as a real physical thing, but rather a representation of the observer's knowledge.  Thus, when an observation is made, at the moment the information about the particle position reaches the cognition of the observer, the wave function must instantaneously collapse from something having infinite extent down to a delta function at some point.  

Because particles are believed to have no representation outside of the wave function under the Copenhagen interpretation, and the wave function is purely a function of the observer's knowledge, this theory implies that there is no objective manifestation of reality-- in other words, that nothing really exists outside of the mind of the observer.

Some common objections to the Copenhagen interpretation include the following: 

\begin{enumerate}
\item The core concept of ``observers'', whose observations determine exactly \emph{when} the wave function collapses, have no definition in the theory beyond the weak implication that any proponent of the theory should consider themself an observer.  As exemplified by the famous ``Schr\"{o}dinger's cat'' thought experiment, in which the cat itself is not considered to be an observer, it is not clear which conscious beings should be considered ``observers.'' Thus, the theory is ambiguous and ill-defined, giving no clear prediction in any situations where the definition of an observer is called into question.
\item Because the positions of particles are only defined relative to the wave function, which is a representation of the observer's knowledge rather than a physical construct, it would be impossible to produce any constructive definition of ``observer'' based purely on particle configurations without resorting to circular logic.
\item Because this interpretation only describes the probability of obtaining a measurement rather than providing a causal description of events and particle interactions that lead up to a measurement, and under this interpretation it is also believed that there is no deeper explanation beyond these statistics, this interpretation effectively implies that there is simply no causal relationship between particle interactions at the quantum level, and hence any attempt to determine such interactions by simulation would be fundamentally misguided.  This results in a lack of ability to describe the dynamical interactions of many particle systems, which would not be possible if all particles existed in indeterminate states.
\item It requires a concept of ``superposition'', whereby a particle's position may be equal to an infinite number of different values simultaneously, which is normally considered the definition of a logical contradiction.
\item Because it relies on fundamental randomness, but no positive definition for randomness can be constructed axiomatically, this implies that the theory can never be fully defined using a strict language like mathematics, suggesting that the interpretation is fundamentally flawed at the logical level.
\end{enumerate}

These vagaries and apparent contradictions, combined with the inability to unify quantum mechanics, general relativity and inflation under a single theory, have motivated many to search for more causal explanations of quantum mechanics.

One of the first such alternatives was the ``theory of the double solution" proposed by \citet{deBroglie1927,deBroglie1970}, in which it was suggested that particles are associated with waves of two different scales: a low-frequency wave associated with some ``internal clock'' of the particle, and a high-frequency wave that he termed the ``pilot wave'' that guides the motion of the particle according to the gradient of the wave (termed the ``guiding equation''), with the time-evolution of both waves governed by the relativistic, real-valued Klein-Gordon equation (which may be derived from the relativistic energy-momentum relation) \citep{Schwabl1999}.

However, as de Broglie was lacking a physical mechanism to explain the production of the double waves \citep{Dagan20}, the theory fell out of favor and became overshadowed by the simpler, single pilot wave model proposed by Bohm \citep{Bohm1952}, whose time evolution was governed by the non-relativistic, complex-valued Schr\"{o}dinger equation.  This theory is now commonly known as Bohmian Mechanics \citep{Goldstein21}.

It has been shown that if one simply assumes that any particle measurement has normally distributed uncertainty, then propagation of that uncertainty through the  Schr\"{o}dinger equation and guiding equation results in a particle distribution following the Born identity \citep{Goldstein21}.  Thus, simulations of particle trajectories computed using Bohmian mechanics result in particle distributions that agree with the statistical predictions of the Copenhagen interpretation under all cases, including the double-slit experiment (\figref{fig_bohmian_trajectories}) and even quantum entanglement \citep{Hugget09}.

\begin{figure}[H]
\centering
\includegraphics[width=0.5 \textwidth]{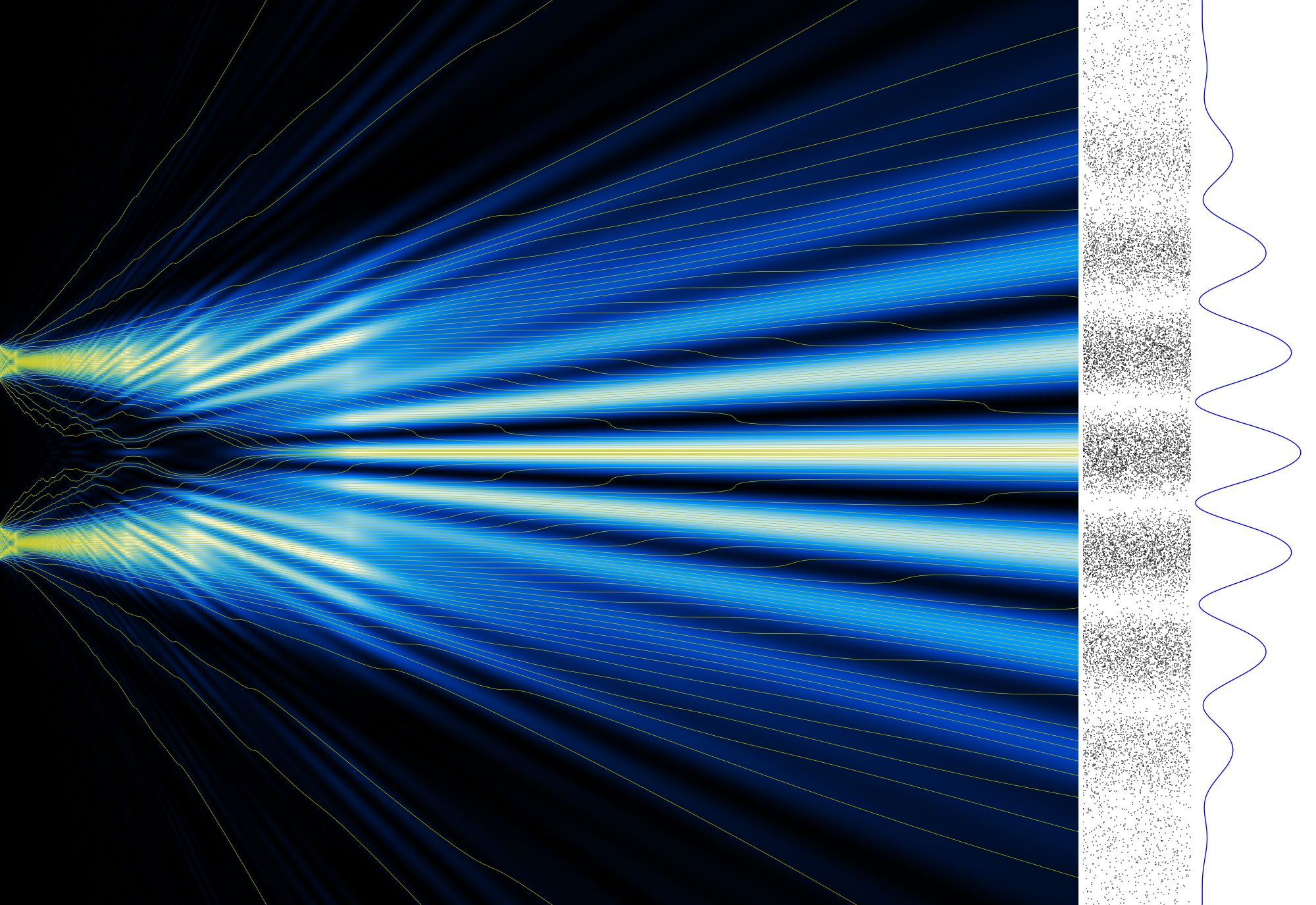}
\caption{Numerical simulation of 100 trajectories guided by the wave function in De Broglie-Bohm's theory by \citet{Gondran07}. Licensed under \protect\href{https://creativecommons.org/licenses/by-sa/4.0/}{CC BY-SA 4.0}.}
  \label{fig_bohmian_trajectories}
\end{figure}

The reverse cannot be said to be true, however, because while Bohmian mechanics is a deterministic set of equations for simulating the time evolution of a quantum system, the Copenhagen interpretation only provides equations for directly quantifying the statistical probability of observing certain outcomes from simple experiments.  As such, the Copenhagen interpretation does not permit one to make predictions when it comes to the time-evolution of quantum systems, systems lacking observers, or multi-particle systems containing complex interactions.

While Bohmian mechanics has proven to be a powerful tool for making quantum scale predictions, it provides no explanation for \emph{why} the concept of both particles and waves should exist simultaneously, \emph{why} particles can apparently come into and out of existence or transmute from one type to another, \emph{why} the wave function should evolve according to a wave equation despite not being recognized as a real fluid, \emph{why} fundamental particles should be associated with the production of waves, or \emph{why} the trajectories of those particles should be guided by the waves according to the gradient of the wave function.  

While it is not necessary for a theory to provide answers to every ``why'' question, and indeed it can be proven that the true equations of physics must rest upon some set of axioms that have no bottom-up explanation \citep{Heinrich12} (other than the strong anthropic principle and the maximally biophilic principle \citep{Heinrich13}), every unanswered question must present itself as an additional axiom which weakens the explanatory power of the theory.   By the law of parsimony, a theory that can make the same predictions from fewer or simpler axioms is more likely to be true, and therefore more deserving of belief.  Thus, one naturally wonders: could there be an even simpler explanation behind Bohmian mechanics?

Along these lines, \citet{Couder06} found that disturbances of a vibrating macroscopic fluid bath may produce droplets of fluid that achieve a stable dynamical state of bouncing in resonance with their own ripples, and that when acted upon by an external force, the interaction of these bouncing droplets with their own ripples results in constant velocity motion, which they called ``inertial walkers.''  After passing through a double-slit, these inertial walkers followed Bohmian-like trajectories and their distribution at a detector followed a wave interference pattern.

These types of 2-dimensional bouncing droplet experiments, now commonly known as Hydrodynamic Quantum Analogs (HQAs) \citep{Bush_2020}, have since been shown to produce many emergent behaviors analogous to behaviors previously thought to be exclusively quantum, such as quantum tunneling \citep{PhysRevLett.102.240401, PhysRevFluids.2.034801, PhysRevE.95.062607}, Landau levels \citep{pnas.1007386107, harris_bush_2014},
Zeeman splitting \citep{Eddi2012}, Friedell oscillations \citep{Tudor2019, doi:10.1126/sciadv.aay9234}, Quantized orbits in a rotating frame \citep{pnas.1007386107, harris_bush_2014,oza_harris_rosales_bush_2014} or a harmonic potential \citep{PhysRevLett.113.104101, Labousse_2014, durey_milewski_2017}, quantum-like statistics in orbital stability \citep{harris_bush_2014,oza_harris_rosales_bush_2014}, effects similar to quantum superposition and the quantum mirage \citep{Saenz2014}, and much more.

Given the astonishing success of HQAs in modeling quantum phenomen, one cannot help but wonder if perhaps quantum mechanics might be the result of similar fluid dynamics of some quantum scale fluid.  If so, this would provide a unified explanation for the existence of both particles and waves under the de Broglie-Bohm pilot wave models, as particles could simply be explained as droplets of fluid held together by surface tension, and waves could be explained as regular mechanical waves produced by the bouncing of droplets on the surface.  Furthermore, HQAs may hint at a deeper explanation for why particles under Bohmian mechanics are guided according to the gradient of the wave, because in HQAs, this occurs as a result of the downward momentum of the bouncing particle being reflected by the surface gradient of the fluid.

Although further experimentation with HQAs by \citet{Anderson15} and \citet{Pucci18} have so-far failed to reproduce the exact interference patterns observed in the quantum double-slit experiment, the trajectories of walkers were confirmed to be influenced by the interference of the carrier wave passing through both slits \citep{Frumkin22}.  The problem, as described by \citet{Anderson15}, is that while the initial carrier wave may travel through both slits, the particle (i.e., bouncing droplet) only goes through one slit, and continues producing additional ripples on that path, whereas the ripples that enter through the opposite slit quickly dissipate, becoming too weak to produce a symmetric wave interference pattern.  Thus, it seems that while the current HQA's show surprising similarity to quantum mechanics, there is something missing.

\citet{Dagan20} have noted that the ripples produced in HQAs differ somewhat from those predicted under Bohmian mechanics as a result of the droplets bouncing on the surface, rather than being simply glued to the wave surface, as assumed under Bohmian mechanics.  They showed that the ripples produced by droplets in HQAs were actually better described by the waves in de Broglie's original, relativistic double-wave solution, if one simply assumes that de Broglie's ``internal clock'' results from the bouncing of the droplet, and that the high frequency waves are due to internal vibrations of the droplet that result in ripples in the surface during the contact interval with the fluid surface on each bounce.

Inspired by this observation, they extended de Broglie's double solution \citep{deBroglie1927,deBroglie1970} using the forced Klein-Gordon equation, where a spin-0 particle is modeled as a localized excitation of its pilot wave field.  Their 1-dimensional model, dubbed Hydrodynamic Quantum Field Theory (HQFT) \citep{Dagan20}, was given by

\begin{align}
\pdd{\phi}{t^2} - c^2 \pdd{\phi}{x^2} + \omega_c^2 \phi = - \epsilon_p f(t) \delta_a(x-x_p) \text{,} \label{eq_one_dimensional_KG}
\end{align}

\noindent where $\phi$ is the elevation of the wave surface, $w_c$ is the Compton frequency (corresponding to the frequency of the droplet bouncing), $\epsilon_p$ is a constant, 

\begin{align}
f(t) = \sin( 2 \omega_c t)
\end{align}

\noindent is the high-frequency localized wave that results from internal vibration of the droplet while in contact with the surface (assumed here to occur at twice the Compton frequency), and 

\begin{align}
\delta_a(x) = (1/|a|\sqrt{\pi}) \exp(-(x/a)^2)
\end{align}

\noindent serves to localize the vibration induced wave, with $a = \lambda_c / 2$.  This latter term serves to approximate the intensity based falloff that would occur due to the expansion of the ripples.

\citet{Jamet21} have generalized \eqnref{eq_one_dimensional_KG} to 3-dimensions (with some modifications), where it was used to successfully model the Bohr-Sommerfeld quantization rule for electron orbitals.  However, they did not attribute any particular underlying mechanism to the production of those waves (such as the bouncing of particles or internal particle vibrations, as proposed under HQFT).

We find HQFT compelling for the way in which it unifies de Broglie-Bohm pilot wave models with insights learned from HQAs.  However, it is unclear whether \citet{Dagan20} mean to propose that \emph{real} quantum particles are local excitations of the pilot wave field which are approximated by bouncing droplets of fluid in HQAs, or vice versa, that real quantum particles are droplets of some quantum scale fluid, which they have merely modeled as local excitations of a field for computational reasons.   

Although they make no mention of a real quantum scale fluid, their proposed mechanism for wave production via the walking droplet system seems to suggest the latter interpretation, in which case \eqnref{eq_one_dimensional_KG} would have to be regarded as an approximation of the surface waves of that quantum scale fluid.  

We note that several important questions remain unanswered under HQFT: 

\begin{enumerate}
\item If waves are produced by a bouncing droplet that only travels through one slit, then how can the interference pattern at the detector appear to be produced by symmetric waves?  
\item If droplets are bouncing on the surface of the fluid, then what is the force that causes them to fall ``down'' (towards the surface of the fluid)? 
\item How can the proposed mechanism of wave production be generalized to the usual 3-dimensions of space?  In other words, if the waves are produced by a particle bouncing off a fluid surface, then in what dimension of space does the bouncing occur?
\item If quantum particles are actually bouncing on a fluid surface, then why hasn't anyone been able to observe them actually bouncing?  In other words, why would this bouncing be described as an ``internal clock'' by de Broglie, rather than an obvious external bouncing motion?
\item Given that HQA's require vibration of the fluid bath, what would be the source of the analogous vibrations in a real quantum fluid?
\end{enumerate}

\section{Quantum Fluid Dynamics} \label{sec_QFD}

Our proposed theory of Quantum Fluid Dynamics (QFD) is almost entirely premised on just a single hypothesis: that ``elementary'' particles (such as electrons, photons, etc.) are not elementary, but rather are droplets formed of a real, quantum-scale fluid that exist in stable dynamical states, akin to the ``inertial walkers'' observed by \citet{Couder06} and others in macroscopic Hydrodynamic Quantum Analogs (HQAs) \citep{Bush_2020}.  Everything else that we conclude follows directly from this hypothesis when combined with other experimentally observable facts.

We refer to this real quantum fluid as the \emph{quantum ocean}, or simply the \emph{ocean} for brevity.  In accordance with the dynamics of HQAs, particles are represented by droplets of the ocean fluid that have been disturbed from the surface, and bounce on the surface in stable dynamical states of resonance with their own ripples.  

The bouncing of these droplets results in the production of low-frequency waves at the droplets bouncing frequency, as well as high frequency ripples that result from internal vibrations of the droplet during the contact interval of the droplet with the fluid surface on each bounce.  As such, our model is conceptually similar in premise to Hydrodynamic Quantum Field Theory (HQFT) \citep{Dagan20}, and thus the local surface dynamics of spin-0 particles in 1-dimension under QFD can be approximated by \eqnref{eq_one_dimensional_KG}.  

However, QFD differs from HQFT in several key regards: (1) we generalize to 3-dimensional space; (2) we consider the whole fluid volume rather than just the local surface wave dynamics; (3) we propose to model particles (i.e., bouncing fluid droplets) and the associated waves that they create as purely emergent phenomena that result from the underlying fluid dynamics equations (i.e., the Euler equations, as discussed in \secref{sec_simulation}), rather than explicitly defining particles as idealized points, or their associated waves as idealized wave functions.

Because elementary particles have 3-degrees of freedom in space, and all particles under the QFD hypothesis are actually small droplets of fluid that are constrained to bouncing on the fluid surface, this implies that the fluid must have a 3-dimensional surface, where the usual 3-dimensions of space actually correspond to the local 3-dimensional tangent space of the ocean surface manifold.  Thus, the embedding space must have an additional degree of freedom, meaning that the ocean itself must exist in a 4-dimensional space.

Because photons are also particles that would be constrained to bouncing on this fluid surface, it would be impossible to observe (with photons, or any other particle) any motion that occurs orthogonal to the surface.  Thus, any bouncing of particles off the surface into the 4th dimension would be invisible (\figref{fig:bouncing_visualization}), and hence the only way detect this bouncing motion would be indirectly from the ripples that are produced.

\begin{figure}[H]
\centering
\includegraphics[width=0.5 \textwidth]{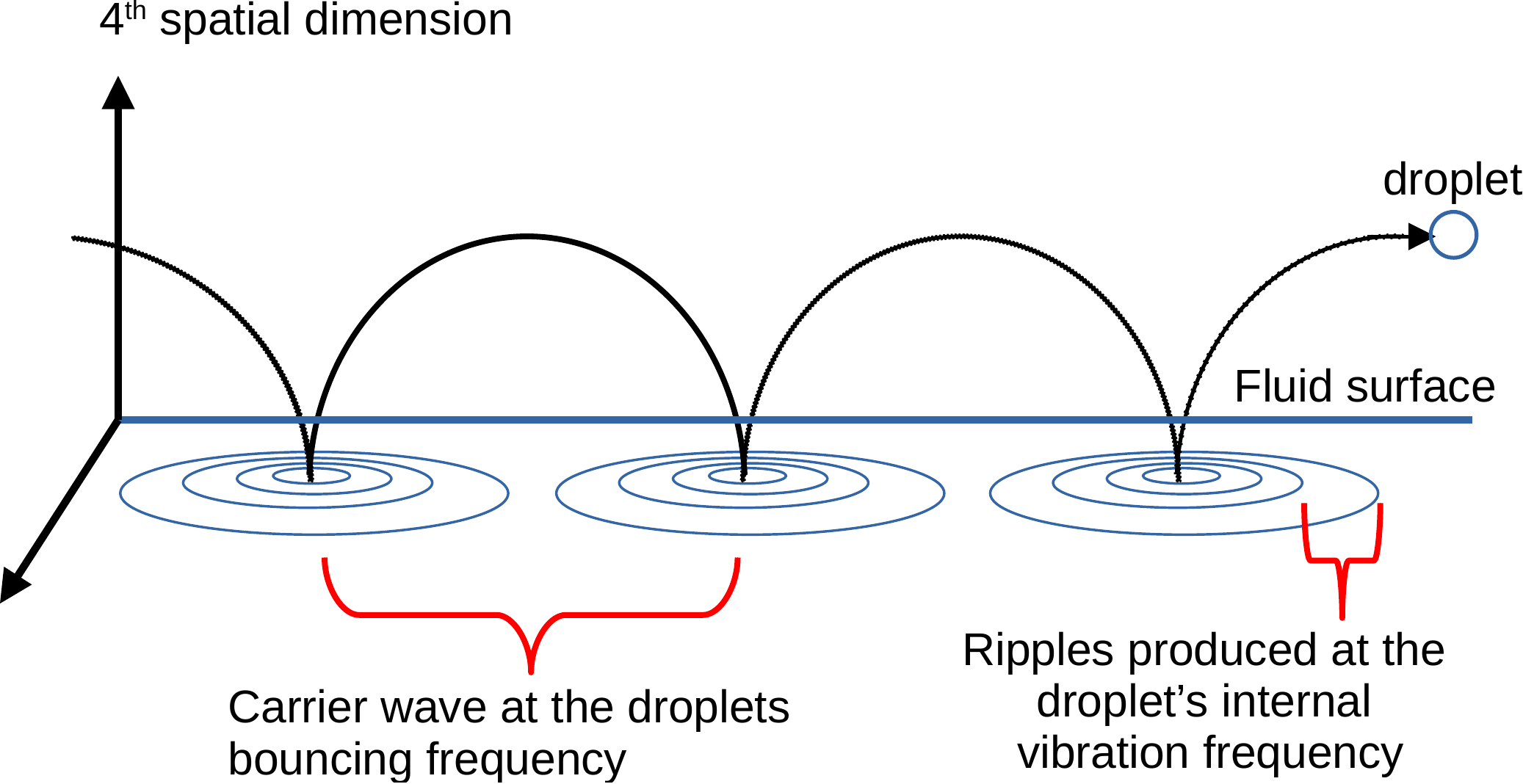}
\caption{Droplets may achieve resonance with the large scale ripples produced by their bouncing motion, resulting in linear motion.  On each bounce, internal droplet vibrations result in higher frequency oscillation while in contact with the surface.}
  \label{fig:bouncing_visualization}
\end{figure}

Because individual particles (such as photons, electrons, etc.) are able to exist indefinitely without sign of dissipation, this implies that the bouncing of droplets must occur without any loss of kinetic energy.  Thus, the quantum ocean must be a superfluid.  Superfluids, as observed for example in certain isotopes of helium at very cold temperatures, have zero viscosity and allow flows without any loss of kinetic energy.  This can explain why bouncing droplets, having achieved resonance, would be capable of bouncing indefinitely.

Because the ocean fluid must able to form droplets and propagate mechanical waves, it must have surface tension, and surface tension results from internal cohesive forces.  Thus, the quantum ocean must be composed of some fluid particle(s) subject to both a mutually attractive cohesive force, which we refer to as \emph{quantum cohesion}, and a shorter range repulsive force that prevents the cohesive force from collapsing all fluid particles into a singularity, which we call \emph{quantum repulsion}.  These forces are analogous to the Van der Waals force of a conventional fluid in HQAs, which switches from an attractive force to a repulsive force at the Van der Waals contact distance \citep{DZYALOSHINSKII1992443}.

Because QFD attempts to explain all known particles as being droplets of this fluid in various stable dynamical states, there is no reason to assume the presence of any other external forces beyond the internal cohesive and repulsive forces of the fluid itself.  Thus, under the dominant influence of the longer-range mutually attractive cohesive force, the overall shape of the ocean must be a sphere, which in a 4-dimensional space would be a hypersphere.

Understanding that the ocean must form a 4-dimensional hypersphere, it also becomes clear that the ``downward'' force acting upon fluid droplets enabling their bouncing, analogous to the force of gravity in HQAs, must simply be the force of quantum cohesion.  In other words, quantum cohesion would attract any droplet that was disturbed from the ocean hypersphere back down towards the surface analogous to the way that gravity attracts mass to the Earth surface.  

This would also explain the mechanism for the gradient based ``guiding equation'' in de Broglie-Bohm pilot wave models as a reflection of the downward cohesive force based on the surface angle of the ocean surface.

As for the double-slit experiment, we believe that the answer may be a combination of two factors.  First, superfluidity means that incident waves will not rapidly dissipate after traveling through either side of the slit, as was identified to be one of the core problems by \citet{Anderson15}.  Secondly, we believe it may be an issue of scale.  Previous experiments with HQA's have all been limited to very rapid periods for the bouncing droplets, assuming that the vertical bouncing motion is fast relative to the horizontal motion \citep{oza_rosales_bush_2013}.  This results in high-frequency ripples being produced almost constantly from the particle, exacerbating the asymmetry due to the particle traveling through one slit.  However, we note that in the limit as the bounces become increasingly spaced out, the wave interference pattern incident on the double-slit approaches that of a single circular wave, which results in a symmetric interference pattern (\figref{fig_double_slit}).

\subsection{Mass-energy equivalence} \label{sec_mass_energy_equivalence}

Consider a stationary droplet of fluid that bounces on/off the ocean surface at light speed, having frequency $f$.  All of the usual relations between energy, momentum, wavelength, and frequency are assumed to hold true.  Therefore, the energy associated with this droplet is given by

\begin{align}
E = \hbar f = p c \text{,}
\end{align}

\noindent where $p$ is momentum (entirely orthogonal to the manifold surface), $c$ is the speed of light, and $\hbar$ is the reduced Planck constant (which may be related to surface tension\citep{Dagan20}).

Rearranging, momentum is given by

\begin{align}
p = \hbar f / c \text{,}
\end{align}

\noindent and relativistic mass is defined by

\begin{align}
p = m_{rel} v \text{.}
\end{align}

Using velocity $v = c$, we obtain relativistic mass of the droplet of

\begin{align}
m_{rel} = p/c = \hbar f / c^2 \text{.} \label{eq_relmass}
\end{align}

Note that because the motion of the droplet occurs entirely orthogonal to the manifold surface, the apparent velocity of this droplet from the perspective of any observer whose every means for detection are constrained to the manifold, the droplet would appear to be at rest.  Therefore, from the perspective of an observer constrained to the manifold,

\begin{align}
m_{rel} = m_0 \text{.} \label{eq_rel_is_rest}
\end{align}

In other words, relativistic mass $m_{rel}$ is \emph{perceived} as a rest mass $m_0$ by an observer.

Rearranging \eqnref{eq_relmass} and substituting in \eqnref{eq_rel_is_rest}, we obtain a relationship between apparent rest mass and frequency, given by

\begin{align}
f = m_0 c^2 / \hbar \text{.} \label{eq_compton_frequency}
\end{align}

Note that \eqnref{eq_compton_frequency} is the definition of Compton frequency, which \citet{deBroglie1924,deBroglie1970} so presciently predicted represented an ``internal clock'' that was associated with all massive particles in his pilot wave model of quantum mechanics (a discovery that he was awarded the Nobel prize for).

Because this bouncing motion occurs orthogonal to the surface of the hypersphere, and photons are also constrained by the surface of the hypersphere, QFD predicts that this bouncing frequency would indeed \emph{appear} to be an invisible internal clock as predicted by de Broglie, despite that it is actually an external clock driven by periodic bouncing.

In addition, it is also interesting to note that this bouncing, being a closed and periodic motion in an invisible 4th spatial dimension, may explain the key hypothesis of Kaluza-Klein theory \citep{KLEIN1926}.

To summarize, under QFD, rest mass is an illusion that results from our inability to directly observe the bouncing motion of particles in the non-observable 4th dimension.

This provides a rational explanation for Einstein's famous mass-energy equivalence,

\begin{align}
E = m_0 c^2 \text{,}
\end{align}

\noindent because under QFD, the conversion of rest mass into energy (motion) is nothing other than motion shifting between the 3-observable dimensions of the manifold, and the 4th non-observable dimension.  Thus, the very concept of energy itself is unified under a single meaning: \emph{energy is motion}.

Accordingly, we now show that it is simple to derive Einstein's generalized relativistic energy-momentum relation by considering motion in this extra dimension.

We start from the usual energy-momentum relation, $E = pc$, where $p$ is a scalar representing the total momentum in all dimensions of motion.  In the 4-dimensional space of QFD, this equation may be equivalently written as

\begin{align}
E = \sqrt{ (pc)^2 + (p'c)^2 } \text{,} \label{energy_momentum_4}
\end{align}

\noindent where $p$ represents momentum in the usual 3-dimensions of space defined by the manifold of the ocean, and $p'$ represents the momentum in the non-observable dimension orthogonal to the manifold (\figref{fig:pythagorean_relation}).

\begin{figure}[H]
\centering
\includegraphics[width=0.25 \textwidth]{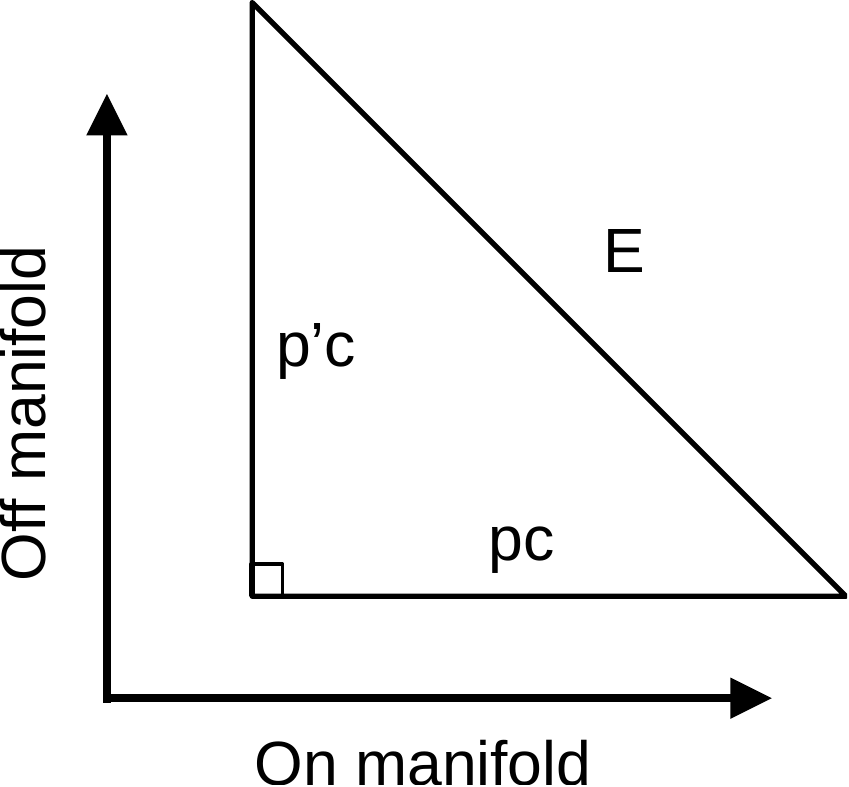}
\caption{Energy-momentum relation in multiple dimensions.}
  \label{fig:pythagorean_relation}
\end{figure}

Although the momentum $p'$ is not directly observable, from \eqnref{eq_relmass}, we know that this off-manifold momentum results in the illusion of rest mass equal to $m_0 = p' / c$.  Rearranging, we find that an observed rest mass $m_0$ implies $p' = m_0 c$.  Plugging into \eqnref{energy_momentum_4}, we obtain

\begin{align}
E &= \sqrt{ (pc)^2 + ( (m_0 c) c)^2 } \\
 &= \sqrt{ (pc)^2 + m_0^2 c^4 } \text{.}
\end{align}

\noindent which of course, is Einstein's relativistic energy-momentum relation.

\subsection{Gravitation} \label{sec_gravitation}

So far, we have shown that particles that bounce on/off the 3-dimensional manifold of the ocean surface have non-observable motion, resulting in the illusion of rest mass from an observer's frame of reference on the manifold.

To create the illusion of mass means that any phenomenon thought to depend on mass must actually have a deeper explanation that can be explained through QFD as a result of bouncing on/off the manifold.  The primary phenomenon that depends on mass is gravitation, the mutual attraction between masses, and the associated warping of space that influences the motion of non-massive particles, such as photons.

In the Newtonian limit, this mutual attraction is approximated by the apparent force of gravity,

\begin{align}
F_g = \frac{G m_1 m_2 }{r^2} \text{,} \label{eq_gravity}
\end{align}

\noindent where $m_1$ and $m_2$ are two rest masses, and $r$ is the distance between them.

Under QFD, we have already shown that rest mass is an illusionary property that is proportional to the frequency of off-manifold bouncing, as defined by \eqnref{eq_relmass}.  

Therefore, plugging \eqnref{eq_relmass} into \eqnref{eq_gravity}, 

\begin{align}
F_g &= \frac{(G \hbar^2 /c_4) f_1 f_2 }{r^2} \\
&= \frac{ G' f_1 f_2 }{r^2} \text{,} \label{eq_gravity2}
\end{align}

\noindent we see that \eqnref{eq_gravity2} describes a long-range effective attraction between bouncing droplets along the manifold that is proportional to the droplet bouncing frequencies, where $r$ is the distance between them on the manifold and $G'$ is a revised gravitational constant.

We now show that this long range attraction is naturally predicted from the quantum fluid dynamics of bouncing droplets.  

First, we note that each time that a droplet bounces off the fluid surface, two principal effects occur:

\begin{enumerate}
\item During the brief interval of time in which the droplet is in contact with the fluid surface, internal vibrations of the droplet produce a burst of circular waves with frequencies corresponding to the internal vibrational frequencies of the droplet.  As these circular waves expand across the hypersphere surface, the amplitude of the waves is reduced according to an inverse square relation.  These ripples interact constructively and destructively with the ripples produced by every other particle in the system.  Because these ripples are zero mean, in the limit as the number of particles goes to infinity, with each particle emitting circular waves from a different position and having uncorrelated phase, the net interference pattern would cancel out and produce no long-term contribution.  Therefore, their contribution to the long range force of gravitation may be ignored.

\item According to conservation of momentum, some downward momentum from the droplet particle must be transferred from the droplet into the fluid surface.  This can be equivalently confirmed by considering that the droplet spends the majority of its time \emph{above} the fluid surface, and so therefore, the surface must exhibit an equal and opposite depression in the time average.
\end{enumerate}

While it is easy to see that the momentum of the bouncing particles should produce some deflection of the surface, it is not immediately apparent why such a deflection would be proportional to frequency.  Indeed, if one assumes that the bouncing droplet is in a state of resonance with the deflection of the surface, as would be the case for example with a ball bouncing on a trampoline, then the frequency of bouncing should have no impact: since both the droplet and the surface would be following sinusoidal paths, the time average position of the surface would be a constant deflection regardless of frequency.

On the other hand, if the surface were being impacted by a force at a frequency that is greater than the ability of the surface to recover from the deflection, for example by being impacted by a jackhammer that is resting on the surface, then it is easy to see how the frequency of such impacts would be proportional to the total deflection.  However, this is only possible because a jackhammer reciprocates by its own action, whereas the energy imparted to the bouncing droplet must come from the medium it is bouncing on.  

In order for a bouncing droplet to bounce at a faster frequency than the surface deflection can recover requires some additional source of upward momentum to be added to the droplet on each bounce.  Cosmological inflation provides exactly this.  The existence of cosmological inflation, as approximated by Hubble's law and later confirmed to be accelerating according to the Friedmann equations \citep{Friedman1922}, implies that the radius of the ocean hypersphere must be increasing at an accelerating rate, and this expansion of the surface would therefore provide additional upward momentum to the bouncing droplet.  In other words, the situation of the bouncing droplet is more akin to jumping on a trampoline while it is accelerating upwards: the upward velocity of the trampoline provides the energy needed to enable the droplet frequency to increase beyond the resonant frequency of the surface.  

Therefore, we conclude that the maximum depression $D(f)$ of the surface from a single droplet should be proportional to frequency,

\begin{align}
D(f) \propto f \text{.} \label{depression_related_to_f}
\end{align}

The depression must be radially symmetric, and the falloff must be inversely related to distance as the force is spread over an increasing area and propagated outwards due to surface tension and internal spring forces of the fluid.  Due to superfluidity, the range of the depression may be infinite.  

The exact rate of falloff would presumably be a function of surface tension and other internal forces of the quantum ocean that escape our present knowledge.  Therefore, without a better understanding of the properties of this quantum superfluid, we have no way to calculate the exact falloff profile in a bottom-up fashion.

However, given the known inverse square relationship to distance in \eqnref{eq_gravity2}, the QFD hypothesis implies that the falloff profile must satisfy

\begin{align}
\lim_{r \rightarrow \infty} y(r) \propto -1/r  \text{.} \label{eq_profile_constraint}
\end{align}

This can be demonstrated using a simple profile of 

\begin{align}
y(r) = \frac{-D(f) r_d}{r}  \text{,} \label{eq_gravity_well}
\end{align}

\noindent where $r_d$ is the radius of the droplet, $D$ is the maximum depression at the droplet, and $r$ is radial distance (\figref{fig:gravitational_well}).  

\begin{figure}[H]
\centering
\includegraphics[width=0.5 \textwidth]{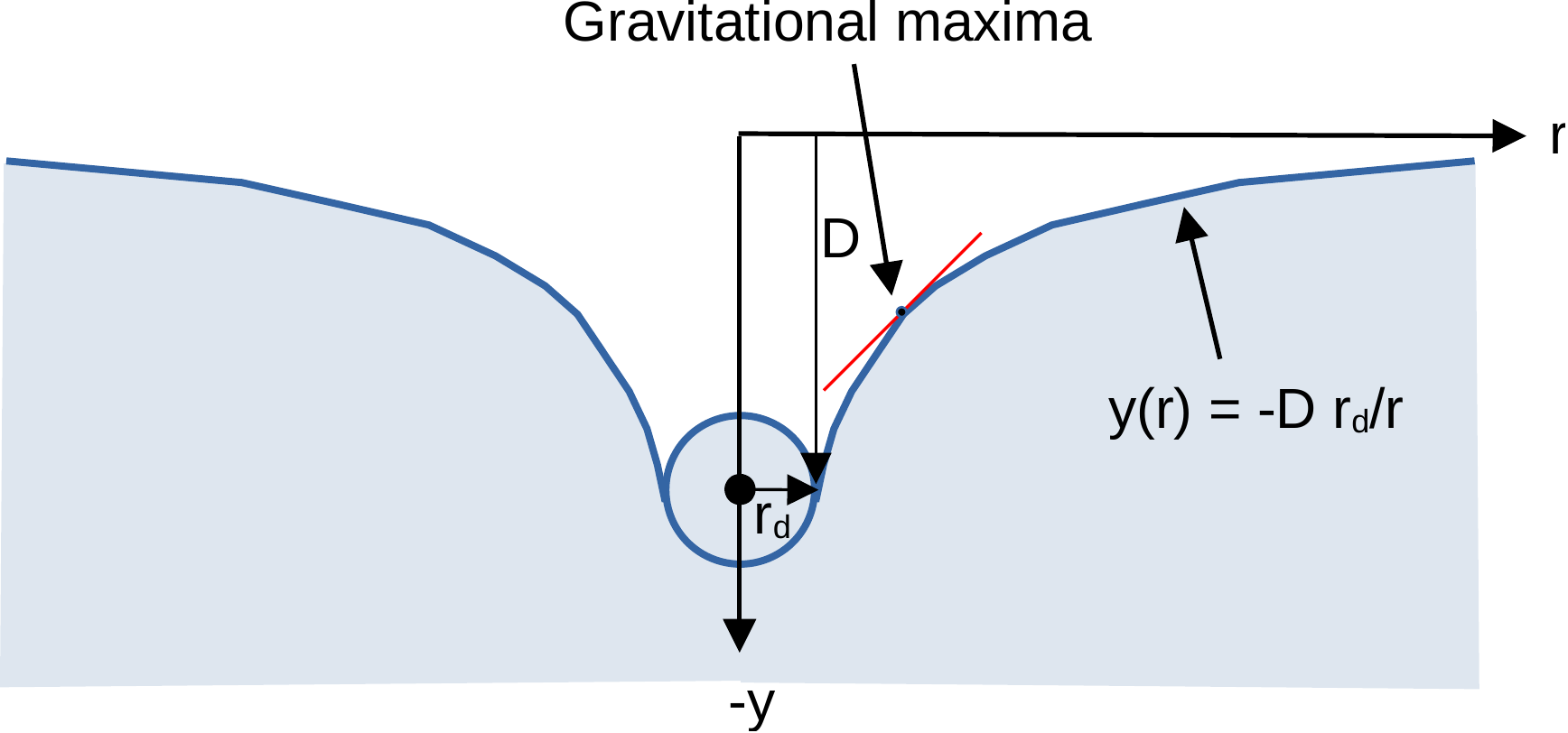}
\caption{Simplified deflection model of the hypersphere surface as a function of radius surrounding a bouncing droplet.}
  \label{fig:gravitational_well}
\end{figure}

Taking the derivative of \eqnref{eq_gravity_well} shows that the slope of the tangent line has an inverse square relation,

\begin{align}
\frac{d}{dr} y(r) \propto \frac{1}{r^2} \text{.}
\end{align}

From simple trigonometry (\figref{fig:bouncing_angle}), this would imply a surface angle of 

\begin{align}
\theta \propto \atan( 1/r^2 ) \text{,}
\end{align}

\noindent and that the proportion of downward momentum that is reflected across the manifold towards the attractor is given by

\begin{align}
F(r) \propto \sin( \pi - 2 \atan(1/r^2) ) \text{.} \label{eq_sin_gravity}
\end{align}

We now investigate in the limit as $r$ tends to infinity.  First, we note that $\lim_{x \rightarrow \infty} 1/x^2 = 0$, and $\lim_{x \rightarrow 0} \atan(x) = x$.  Therefore,

\begin{align}
\lim_{r \rightarrow \infty} F(r) \propto \sin( \pi - 2/r^2 ) \text{.}
\end{align}

Furthermore, $\lim_{x \rightarrow 0} \sin(\pi - x) = -x$.  Therefore,

\begin{align}
\lim_{r \rightarrow \infty} F(r) \propto 1/r^2 \text{.}
\end{align}

\begin{figure}[H]
\centering
\includegraphics[width=0.5 \textwidth]{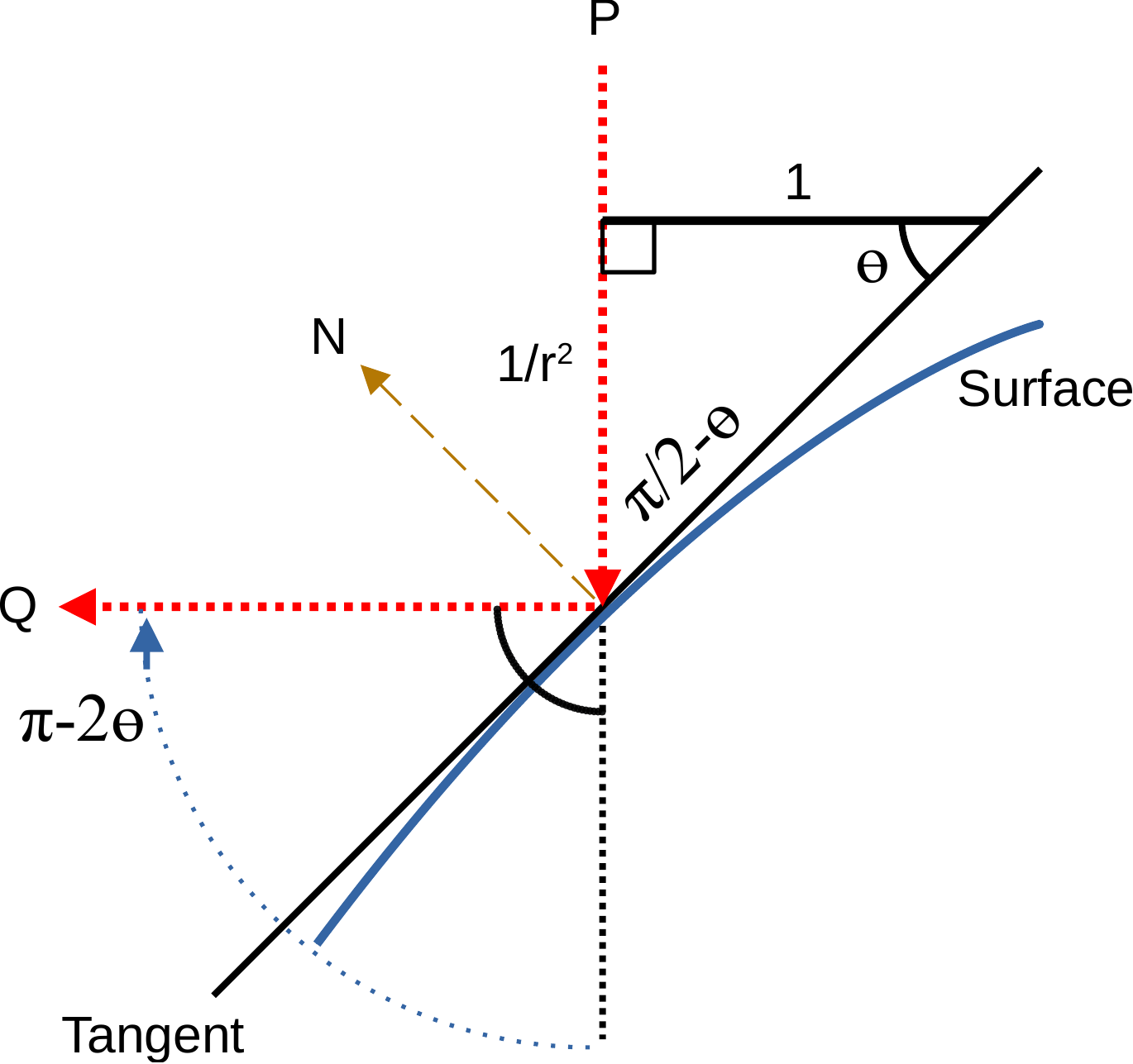}
\caption{Reflection of droplet with vector $P$ is reflected to vector $Q$ about the surface normal $N$.  The slope of the local surface tangent is $1/r^2$, resulting in a surface angle of $\theta = \atan(1/r^2)$.   The reflection vector has angle $\pi - 2 \theta = \pi - 2 \atan(1/r^2)$ with respect to $P$. }
  \label{fig:bouncing_angle}
\end{figure}

Thus, we see that any falloff profile satisfying \eqnref{eq_profile_constraint} would result in an apparent force of gravity that is proportional to the inverse square of the distance between masses, as predicted in the Newtonian limit  (\figref{fig:force_emergence}).  This profile constraint may imply some characteristics on the internal forces of the quantum fluid that merit further study.

Because the surface deflection resulting from multiple bouncing particles in close relative proximity is always in the same direction (towards the center of the hypersphere), the long-range depression of the surface from objects composed of many particles would add together constructively, meaning that the effective net force between large massive bodies (such as planets, stars, etc.) would be well-predicted by the same equation using the sum of the constituent particle masses.

Furthermore, because all particles are constrained to bouncing on/off the manifold, these gravitational wells would also influence any non-massive particles, causing them to follow the geodesic equations of motion as predicted by general relativity.  Therefore, under QFD, the warping of ``spacetime'' under general relativity is really just warping of the ocean surface.

\begin{figure}[H]
\centering
\includegraphics[width=0.5 \textwidth]{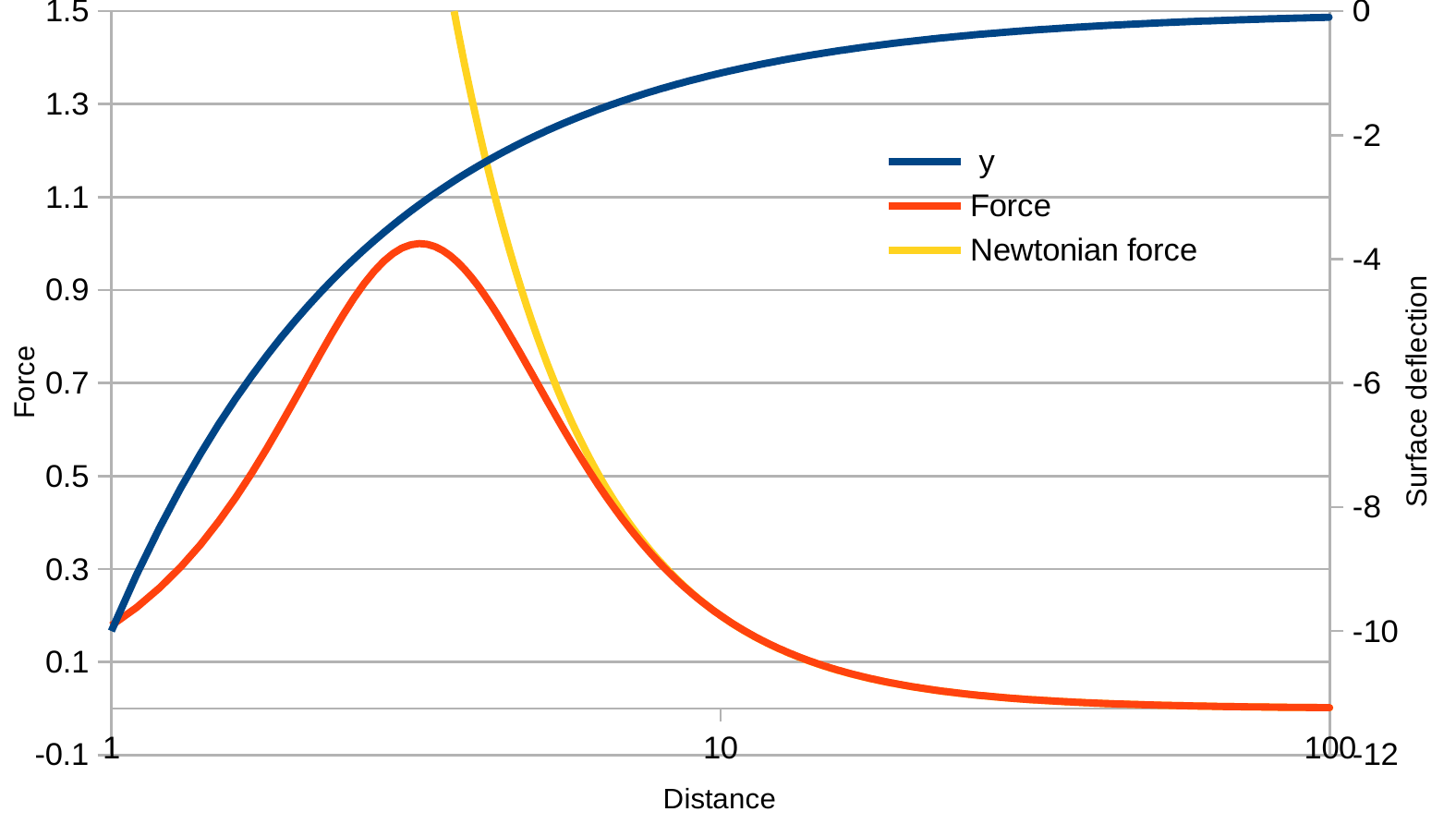}
\caption{Predicted effective gravitational force compared to Newtonian force of gravity, plotted on a logarithmic distance scale.  Gravitational well defined by a droplet of radius $r_d = 1$ and maximum deflection of $D = 10$.  The maximum attractive force occurs at $r \approx 3.2$.  For distances greater than about five times the droplet radius, the predicted force quickly converges to an inverse square relation, consistent with Newtonian gravity and general relativity.}
  \label{fig:force_emergence}
\end{figure}

In summary, QFD predicts that massive bodies would warp the 3-dimensional surface manifold of the ocean (which corresponds to the perceived 3-dimensions of space), resulting in the appearance of a long-range, mutually attractive apparent force that is proportional to the product of both apparent masses, and would be inversely related to the squared distance between masses if the depression is consistent with \eqnref{eq_profile_constraint}.

Finally, we note an interesting deviation from the predictions of Newtonian gravity (and general relativity) in the limit as distance between two elementary particles goes to zero.  Whereas Newtonian gravity predicts infinite gravitational force, and general relativity predicts a divergence of spacetime curvature with no solution, QFD instead predicts that effective gravitational attraction achieves a maximum when the slope of the gravitational well reaches a 45 degree angle.  At this angle, the droplet will be reflected 90 degrees from it's original downward trajectory (and may bounce off the opposing side of the gravitational well).  In other words, at extremely close distances (on the order of the droplet radius), a bouncing droplet may become physically trapped in the gravitational well.  However, at this distance the droplets would also have a high probability of physically intersecting, in which case the droplets might merge or change their dynamics (i.e., transmuting into different particle(s)).

\subsection{Inertia and Momentum} \label{sec_inertia_and_momentum}

Because inertia and momentum are fundamentally related to particle mass, which under QFD is merely an emergent phenomenon resulting from droplet bouncing frequency (\secref{sec_mass_energy_equivalence}), we now show inertia and momentum can be derived from QFD.

The principle of inertia states that any object at rest or in motion will maintain constant velocity until acted upon by an external force.  The force required is formalized by

\begin{align}
F = m a \text{,} \label{eq_fma}
\end{align}

\noindent where $F$ is the force required to produce an acceleration of $a$ for an object of mass $m$.

Bouncing droplets must always be in resonance with the low frequency component of their ripples, because these ripples are produced from their own bouncing.  As demonstrated by experiments with HQAs, this resonance causes droplets that are influenced by an external force to become propelled by the guidance of their own ripples, which effectively holds their velocity constant until acted upon by another external force\citep{Dagan20}.

Thus, any gradual change in velocity would require overcoming the guiding force of the waves that are actively holding velocity constant.  This explains why the force required to change the velocity of a bouncing droplet must be proportional to the acceleration in \eqnref{eq_fma}.

The additional proportionality to mass can be explained naturally as a result of the fact that bouncing particles bury themselves in the surface of the manifold by creating gravitational wells \eqnref{eq_gravity_well} with a maximum depression that is proportional to their bouncing frequency \eqnref{depression_related_to_f} which is perceived as rest mass \eqnref{eq_compton_frequency}.

It is now straight-forward to show how mass is related to momentum.  By definition, $\frac{d}{dt} p(t) = F(t)$ and $\frac{d}{dt} v(t) = a(t)$.  Substituting into \eqnref{eq_fma}, 

\begin{align}
\frac{d}{dt} p(t) = m \frac{d}{dt} v(t) \text{,}
\end{align}

\noindent and taking the indefinite integral of both sides,

\begin{align}
\int \frac{d}{dt} p(t) dt  &= m \int \frac{d}{dt} v(t)  dt \\
p(t) &= m v(t) \\
p &= m v
\text{.}
\end{align}

\subsection{Quantum fluctuations} \label{sec_fluctuations}

Due to the superfluid nature of the ocean, ripples produced by bouncing droplets anywhere on the hypersphere surface (i.e., anywhere in the universe) would propagate infinitely, causing endless and unpredictable background fluctuations that depend, to some degree, on the complete history of every other particle in the entire universe.

This provides a coherent explanation for the observational existence of quantum vacuum fluctuations \citep{Yu2020} and of Heisenberg's uncertainty principle \citep{Debashis14}: considering that there is fundamentally no way to directly measure the turbulent state of the fluid surface surrounding the gross particle positions in any quantum experiment, it would be fundamentally impossible to ever measure the exact initial state of any local quantum system.

These unpredictable background fluctuations are analogous to the vibration of the fluid bath in HQA experiments.

\subsection{Massless particles} \label{sec_massless}

Under the QFD hypothesis, all particles are represented by droplets of fluid that bounce on the hypersphere surface.  In \secref{sec_mass_energy_equivalence}, we showed that these particles are perceived to have rest mass proportional to their bouncing frequency (the Compton frequency).  But what about massless particles, such as photons?

As discussed in \secref{sec_gravitation}, gravitational wells only occur in QFD as a result of having bouncing frequencies greater than the maximum restitution frequency of the fluid surface, which is made possible by the extra upward momentum provided by inflation.  Therefore, for particles that have bouncing frequencies less than the maximum restitution frequency of the fluid surface, the bouncing of the droplet will be in resonance with the fluid surface and thus the proportional relationship between frequency and maximum surface deflection (and hence apparent rest mass) would disappear.

For such particles, there would only be a very miniscule depression to cancel out the momentum of the droplet, independent of frequency.  It is likely that the scale of this depression would be so small that it is less than the scale of the quantum background fluctuations (\secref{sec_fluctuations}), in which case it would be perceived as a totally massless particle.

\subsection{Inflation and time} \label{sec_inflation_and_time}

Based on cosmological observations, it is known that space is undergoing uniform inflation as approximated by Hubble's law, and that this inflation is occurring at an accelerating pace according to the Friedmann equations \citep{Friedman1922}.  

In the context of the QFD, cosmological inflation implies that the radius of the ocean hypersphere is expanding.  We note that this expanding radius could potentially have a fluid dynamics explanation.  For example, it could be that the cohesion force is weakening over time, or that the repulsive force is strengthening over time.

This brings us to the subject of time, which is interesting because the 3-dimensional surface manifold of the ocean hypersphere (which, under QFD, corresponds to the perceived 3-dimensions of space) must be continually moving outward in the 4th dimension (orthogonal to the manifold), and this 4th dimensional motion is positively associated with the forward progression of time.  Thus, from a practical perspective, the 4th spatial dimension behaves \emph{as if} it were a time-like dimension, despite being a regular dimension of space.

In addition, we showed that the bouncing frequency of particles (which is perceived as rest mass) results in deformation of the surface by creating gravitational wells that offer the only possible means for particles to move through this 4th dimension (other than the natural progression due to inflation).  Thus, we see that there is an interesting relationship here with mass deforming the regular spatial dimensions through the 4th dimension that is uniquely associated with time.

Furthermore, we note that curvature of the surface in this 4th dimension affects the rate of bouncing of all droplets (i.e., matter particles).  When a bouncing droplet approaches a gravitational well, the rate of bouncing must effectively slow down, because the height of the surface at each successive bounce is lower than the previous.  Thus, the ``internal clock'' of the particle effectively slows down, which means that for any process that depends on this frequency, the progression of time may appear to slow down.

Thus, one may naturally ask several questions: is this 4th dimension the time dimension?   Does expansion of the hypersphere define the ``arrow of time''?  When droplets bounce on the manifold in the invisible 4th dimension, are they actually bouncing through time?  Do gravitational wells, which warp the manifold and allow particles to travel to a lower radius that corresponds to an earlier time, actually represent a motion backwards in time?

In the context of QFD, the answer to all of these questions must be no.  We can say this with confidence because, while inflation may be responsible for gravitation, it is not responsible for enabling all dynamic interactions in the system.  Indeed, our model is premised on the assumption of a dynamical fluid.  One cannot have any dynamics, as exemplified by fluid flow, without a pre-existing concept of change, and one cannot have change without time.  Thus, while the progression of time may be positively associated with movement in the 4th dimension, the concept of time must be much more fundamental.

By definition, each dimension of space provides an additional degree of freedom for movement.  However, one cannot have movement without the ability for change -- and if change is possible, then a record of those changes can always be recorded and visualized on an axis.  However, we do not believe such a visualization of the time axis should be considered to correspond to a literal dimension because it does not provide any additional degrees of freedom for movement within the system.  One does not \emph{move} through time (either forwards or backwards), because time is merely a record of change.

\subsection{Simulation} \label{sec_simulation}
 
Previous pilot wave models of quantum mechanics, such as de Broglie's double solution \citep{deBroglie1927}, Bohmian mechanics \citep{Bohm1952,Bohm1952b}, and more modern variations like HQFT \citep{Dagan20} have all been formulated in terms of differential surface wave equations (e.g., Schr\"{o}dinger's equation, valid for non-relativistic spin-0 particles \citep{griffiths:quantum}; the Klein-Gordon equation, valid for relativistic spin-0 particles; the Dirac equation, valid for relativistic spin-1/2 particles; and the Pauli equation, valid for non-relativistic spin-1/2 particles \citep{Schwabl1999}).

Under QFD, it is predicted that surface wave models may serve as reasonable approximations to modeling individual particle trajectories under idealized conditions.  However, a fundamental problem with this type of approach is that each type of particle would require different equations, making it impossible to reconcile these types of equations into a fully unified set of equations that describe arbitrary particle interactions.  

Indeed, the surface wave equations are abandoned in modern Quantum Field Theory (QFT) \citep{Peskin95}, in which particles are modeled as non-fundamental excited states of their underlying quantum fields.  In this regard, QFT is similar to the proposed QFD, which models particles as excited states of an underlying fluid.  However, QFT lacks a formal mathematical foundation and, according to Haag's theorem, is fundamentally ill-defined \citep{Haag:212242}.

Moreover, we note that such local surface wave based formulations would not be able model in a bottom-up fashion either the gravitation-like attraction of droplets to the fluid surface, or the inflation of the fluid surface that are predicted under QFD, because these are the net effects resulting from the interacting dynamics of all fluid particles in the ocean hypersphere volume.  We believe this may explain why previous attempts at unification of pilot wave models with cosmic scale observations such as gravitation and inflation have so far been unsuccessful.

Under the QFD hypothesis, the \emph{true} laws of physics must be formulated in terms of the internal interactions between all quantum fluid particles in the hypersphere volume.  From this perspective, all known particles (i.e., all bosons and fermions) would be modeled as droplets of the underlying fluid that exist in various stable dynamical states described by their properties (e.g., mass, spin, charge, mass, polarization, etc.).  While the laws of physics would be incredibly simple and elegant to write down from this perspective, such a ``kinetic theory'' approach to simulation would never be remotely computationally practical.

Fortunately, it is not necessary to resort to low-level modeling of individual quantum fluid particles.  From a Computational Fluid Dynamics (CFD) perspective, it is known that there are many equivalent perspectives from which the \emph{flow} of any fluid can be more efficiently modeled at arbitrary grid (or representative particle ensemble) resolutions.  Fundamentally, models of the flow can be formulated either in conservation form or non-conservation form, integral form or partial differential equation form \citep{anderson1995computational}.  We therefore propose that simulations of QFD should choose appropriate model(s) of the flow based on computational demands.

Although we can make few specific claims about the composition or internal forces of the quantum fluid, the success of HQAs at modeling quantum phenomena suggest that the flow of this quantum fluid must be at least well-approximated by macroscopic fluid flows, and hence the Navier-Stokes equations should be valid.  For a superfluid, the Navier-Stokes equations reduce to the Euler equations for inviscid flow \citep{anderson1995computational}.

There are two primary computational challenges that are specific to QFD simulation.  The first challenge is that because QFD depends on the net long-range cohesive force from all quantum fluid particles in the 4-dimensional hypersphere volume, it would not be possible to simulate this behavior in a purely bottom-up fashion without modeling the \emph{entire} hypersphere (i.e., the entire universe).  Thus, simulating gravitational effects in QFD would either require a multi-resolution model, or more simply, an analytical calculation of the net effect of quantum cohesion and repulsion from the entire hypersphere, which could then be translated into the local frame of a bounded simulation.

Taking this approach, we propose that the 4-dimensional compressible Euler equations of QFD be written in conservation form as

\begin{align}
\pd{\rho}{t} + \mathbf{u} \cdot \nabla \rho + \rho \nabla \cdot \mathbf{u} &= 0\\
\pd{ \mathbf{u} }{t} + \mathbf{u} \cdot \nabla \mathbf{u} + \frac{ \nabla p }{ \rho } &= \mathbf{g} \\
\pd{ e}{t} + \mathbf{u} \cdot \nabla e + \frac{ p}{\rho} \nabla \cdot \mathbf{u} &= 0 \text{,}
\end{align}

\noindent where $\rho$ is the fluid density, $p$ is pressure, $\mathbf{u} \in \mathbb{R}^4$ is flow velocity,  $e$ is specific internal energy, $\mathbf{g} \in \mathbb{R}^4$ is the net acceleration due to internal cohesive and repulsive fluid forces, and $t$ is time.

In density form, we propose that $\mathbf{g}$ may be written as

\begin{align}
\mathbf{g}( \mathbf{x}, t) = \iiiint_V \rho(\mathbf{v}) f( \norm{\mathbf{x}-\mathbf{v}},t) \, dV \text{,}
\end{align}

\noindent where $V \subset \mathbb{R}^4$ is the fluid volume, $\mathbf{v}$ is the variable of integration, and $f(r,t)$ is some distance-dependent function approximating the long-range cohesion and short range repulsion between fluid particles resulting from internal forces of the quantum ocean.  We have made it time-dependent as a potential means for modeling inflation, as discussed in \secref{sec_inflation_and_time}.

Perhaps the most challenging aspect of simulation arises from the fact that because particles in QFD correspond to droplets of fluid that bounce on the fluid surface, the fluid-gas-fluid interface between a droplet and the surface on each bounce must be tracked with extreme precision in order to avoid the simulated droplet from being merged into the fluid body as a result of insufficient grid resolution.  Thus, it is likely that special methods for precise interface tracking would need to be developed that are capable of accurately reflecting the conditions under which a droplet would bounce off the surface of the ocean vs. become reabsorbed by it.

\section{Conclusion}

The remarkable success of Hydrodynamic Quantum Analogs (HQAs) at modeling quantum phenomenon as the dynamics of bouncing droplets using real fluids has motivated us to hypothesize the existence of a real quantum scale fluid having similar dynamics.  This hypothesis forms the sole basis of our proposed theory of Quantum Fluid Dynamics (QFD).  

We showed that if there does exist a real quantum fluid analogous to the fluid in HQAs, then this fluid must be a superfluid, and the dynamics of that fluid must take place on the 3-dimensional surface of a 4-dimensional hyperspherical fluid volume.

We showed that the mechanism by which droplets bounce may be naturally explained as a result of internal fluid cohesive forces, and that this bouncing would occur in a non-observable 4th spatial dimensional resulting in the appearance of a property analogous to rest mass from the perspective of any observer.  

We showed that the frequency of bouncing would be consistent with de Broglie's prediction that matter particles be associated with an invisible internal clock at the Compton frequency.  In addition, we showed that Einstein's mass-energy equivalence and relativistic energy-momentum relations can be derived for bouncing droplets directly from this assumption.

We showed that under the influence of cosmological inflation (which could potentially have a fluid dynamics explanation), the dynamics of bouncing droplets would deform the 3-dimensional surface manifold of the fluid hypersphere proportional to the perceived rest mass, creating literal gravitational wells that influence the motion of other bouncing droplet particles in the manner predicted by general relativity.  In addition, we  showed that inertia and momentum relations can be derived for bouncing droplets relative to the perceived rest mass.

We showed that inflation would cause the 4th spatial dimension to be perceived as a time-like dimension, in that this dimension is positively associated with the forward progression of time, and particles generally cannot move backwards through this dimension except in the presence of gravitational wells, where the internal clock would appear to slow down, creating the illusion of moving through time at a slower rate.

If the QFD hypothesis is true, it would explain \emph{why} HQAs form effective quantum analogs, and would explain why the guiding equation in de Broglie-Bohm pilot wave models (which may be viewed as localized fluid surface level approximations to QFD volume calculations) depends on the gradient of the wave function.

Although QFD is not a fully developed theory, and the potential for reconciliation with electric charge, spin, Lorentz contractions and many other properties has not been thoroughly investigated, QFD demonstrates a remarkable potential to unify many quantum scale observations with cosmological scale observations from a new and very parsimonious perspective.

In conclusion, the proposed model of QFD is not just a theory of quantum mechanics -- it is a highly parsimonious framework that may potentially form the basis of a unifying theory of physics at all scales.  If true, it suggests that the cosmic scale is inextricable from the quantum scale, and that concepts such as mass, momentum, inertia, and general relativity are not be fundamental laws of physics, but rather emergent phenomena that result from quantum scale fluid dynamics.

\bibliography{quantum_fluid_dynamics}

\begin{thebibliography}{42}
\providecommand{\natexlab}[1]{#1}
\providecommand{\url}[1]{\texttt{#1}}
\expandafter\ifx\csname urlstyle\endcsname\relax
  \providecommand{\doi}[1]{doi: #1}\else
  \providecommand{\doi}{doi: \begingroup \urlstyle{rm}\Url}\fi

\bibitem[de~Broglie(1927)]{deBroglie1927}
Louis de~Broglie.
\newblock La m{\'e}canique ondulatoire et la structure atomique de la
  mati{\`e}re et du rayonnement.
\newblock \emph{Journal De Physique Et Le Radium}, 8:\penalty0 225--241, 1927.

\bibitem[Bohm(1952{\natexlab{a}})]{Bohm1952}
David Bohm.
\newblock A suggested interpretation of the quantum theory in terms of "hidden"
  variables. i.
\newblock \emph{Phys. Rev.}, 85:\penalty0 166--179, Jan 1952{\natexlab{a}}.
\newblock \doi{10.1103/PhysRev.85.166}.
\newblock URL \url{https://link.aps.org/doi/10.1103/PhysRev.85.166}.

\bibitem[Bohm(1952{\natexlab{b}})]{Bohm1952b}
David Bohm.
\newblock A suggested interpretation of the quantum theory in terms of "hidden"
  variables. ii.
\newblock \emph{Phys. Rev.}, 85:\penalty0 180--193, Jan 1952{\natexlab{b}}.
\newblock \doi{10.1103/PhysRev.85.180}.
\newblock URL \url{https://link.aps.org/doi/10.1103/PhysRev.85.180}.

\bibitem[Bush and Oza(2020)]{Bush_2020}
John W~M Bush and Anand~U Oza.
\newblock Hydrodynamic quantum analogs.
\newblock \emph{Reports on Progress in Physics}, 84\penalty0 (1):\penalty0
  017001, dec 2020.
\newblock \doi{10.1088/1361-6633/abc22c}.
\newblock URL \url{https://doi.org/10.1088/1361-6633/abc22c}.

\bibitem[Griffiths(1995)]{griffiths:quantum}
David Griffiths.
\newblock \emph{Introduction of Quantum Mechanics}.
\newblock Prentice Hall, Inc., 1995.

\bibitem[Omnes(1999)]{Omnes1999}
Roland Omnes.
\newblock \emph{Understanding Quantum Mechanics}.
\newblock Princeton University Press, 1999.
\newblock ISBN 9780691004358.
\newblock URL \url{http://www.jstor.org/stable/j.ctv173f2pm}.

\bibitem[de~Broglie(1970)]{deBroglie1970}
Louis de~Broglie.
\newblock The reinterpretation of wave mechanics.
\newblock \emph{Founations of Physics}, 1\penalty0 (1):\penalty0 5--15, Jun
  1970.
\newblock URL \url{https://link.springer.com/article/10.1007/BF00708650}.

\bibitem[Schwabl(1999)]{Schwabl1999}
Franz Schwabl.
\newblock \emph{Relativistic Wave Equations and their Derivation}, pages
  115--130.
\newblock Springer Berlin Heidelberg, Berlin, Heidelberg, 1999.
\newblock ISBN 978-3-662-03929-8.
\newblock \doi{10.1007/978-3-662-03929-8_5}.
\newblock URL \url{https://doi.org/10.1007/978-3-662-03929-8_5}.

\bibitem[Dagan and Bush(2020)]{Dagan20}
Yuval Dagan and John W.~M. Bush.
\newblock Hydrodynamic quantum field theory: the free particle.
\newblock \emph{Comptes Rendus. M\'ecanique}, 348\penalty0 (6-7):\penalty0
  555--571, 2020.
\newblock \doi{10.5802/crmeca.34}.

\bibitem[Goldstein(2021)]{Goldstein21}
Sheldon Goldstein.
\newblock {Bohmian Mechanics}.
\newblock \emph{Stanford Encyclopedia of Philosophy}, Jun 2021.
\newblock URL \url{https://plato.stanford.edu/entries/qm-bohm/}.

\bibitem[Huggett and Vistarini(2009)]{Hugget09}
Nick Huggett and Tiziana Vistarini.
\newblock Entanglement exchange and bohmian mechanics, 2009.
\newblock URL \url{https://arxiv.org/abs/0905.4036}.

\bibitem[Gondran and Gondran(2005)]{Gondran07}
Michel Gondran and Alexandre Gondran.
\newblock Numerical simulation of the double slit interference with ultracold
  atoms.
\newblock \emph{American Journal of Physics}, 73\penalty0 (6):\penalty0
  507--515, 2005.
\newblock \doi{10.1119/1.1858484}.
\newblock URL \url{https://doi.org/10.1119/1.1858484}.

\bibitem[Heinrich(2012)]{Heinrich12}
Stuart Heinrich.
\newblock The relativity of existence, 2012.

\bibitem[Heinrich(2013)]{Heinrich13}
Stuart Heinrich.
\newblock Physical relativism as an interpretation of existence, 2013.

\bibitem[Couder and Fort(2006)]{Couder06}
Yves Couder and Emmanuel Fort.
\newblock Single-particle diffraction and interference at a macroscopic scale.
\newblock \emph{Phys. Rev. Lett.}, 97:\penalty0 154101, Oct 2006.
\newblock \doi{10.1103/PhysRevLett.97.154101}.
\newblock URL \url{https://link.aps.org/doi/10.1103/PhysRevLett.97.154101}.

\bibitem[Eddi et~al.(2009)Eddi, Fort, Moisy, and
  Couder]{PhysRevLett.102.240401}
A.~Eddi, E.~Fort, F.~Moisy, and Y.~Couder.
\newblock Unpredictable tunneling of a classical wave-particle association.
\newblock \emph{Phys. Rev. Lett.}, 102:\penalty0 240401, Jun 2009.
\newblock \doi{10.1103/PhysRevLett.102.240401}.
\newblock URL \url{https://link.aps.org/doi/10.1103/PhysRevLett.102.240401}.

\bibitem[Nachbin et~al.(2017)Nachbin, Milewski, and
  Bush]{PhysRevFluids.2.034801}
Andr\'e Nachbin, Paul~A. Milewski, and John W.~M. Bush.
\newblock Tunneling with a hydrodynamic pilot-wave model.
\newblock \emph{Phys. Rev. Fluids}, 2:\penalty0 034801, Mar 2017.
\newblock \doi{10.1103/PhysRevFluids.2.034801}.
\newblock URL \url{https://link.aps.org/doi/10.1103/PhysRevFluids.2.034801}.

\bibitem[Hubert et~al.(2017)Hubert, Labousse, and Perrard]{PhysRevE.95.062607}
M.~Hubert, M.~Labousse, and S.~Perrard.
\newblock Self-propulsion and crossing statistics under random initial
  conditions.
\newblock \emph{Phys. Rev. E}, 95:\penalty0 062607, Jun 2017.
\newblock \doi{10.1103/PhysRevE.95.062607}.
\newblock URL \url{https://link.aps.org/doi/10.1103/PhysRevE.95.062607}.

\bibitem[Fort et~al.(2010)Fort, Eddi, Boudaoud, Moukhtar, and
  Couder]{pnas.1007386107}
Emmanuel Fort, Antonin Eddi, Arezki Boudaoud, Julien Moukhtar, and Yves Couder.
\newblock Path-memory induced quantization of classical orbits.
\newblock \emph{Proceedings of the National Academy of Sciences}, 107\penalty0
  (41):\penalty0 17515--17520, 2010.
\newblock \doi{10.1073/pnas.1007386107}.
\newblock URL \url{https://www.pnas.org/doi/abs/10.1073/pnas.1007386107}.

\bibitem[Harris and Bush(2014)]{harris_bush_2014}
Daniel~M. Harris and John W.~M. Bush.
\newblock Droplets walking in a rotating frame: from quantized orbits to
  multimodal statistics.
\newblock \emph{Journal of Fluid Mechanics}, 739:\penalty0 444–464, 2014.
\newblock \doi{10.1017/jfm.2013.627}.

\bibitem[Eddi et~al.(2012)Eddi, Moukhtar, Perrard, Fort, and Couder]{Eddi2012}
A.~Eddi, J.~Moukhtar, S.~Perrard, E.~Fort, and Y.~Couder.
\newblock Level splitting at macroscopic scale.
\newblock \emph{Physical review letters}, 108, 2012.
\newblock \doi{10.1103/PhysRevLett.108.264503}.

\bibitem[Cristea-Platon(2019)]{Tudor2019}
Tudor Cristea-Platon.
\newblock \emph{Hydrodynamic analogues of quantum corrals and Friedel
  oscillations}.
\newblock PhD thesis, Massachusetts Institute of Technology, 2019.

\bibitem[Sáenz et~al.(2020)Sáenz, Cristea-Platon, and
  Bush]{doi:10.1126/sciadv.aay9234}
Pedro~J. Sáenz, Tudor Cristea-Platon, and John W.~M. Bush.
\newblock A hydrodynamic analog of friedel oscillations.
\newblock \emph{Science Advances}, 6\penalty0 (20):\penalty0 eaay9234, 2020.
\newblock \doi{10.1126/sciadv.aay9234}.
\newblock URL \url{https://www.science.org/doi/abs/10.1126/sciadv.aay9234}.

\bibitem[Oza et~al.(2014)Oza, Harris, Rosales, and
  Bush]{oza_harris_rosales_bush_2014}
Anand~U. Oza, Daniel~M. Harris, Rodolfo~R. Rosales, and John W.~M. Bush.
\newblock Pilot-wave dynamics in a rotating frame: on the emergence of orbital
  quantization.
\newblock \emph{Journal of Fluid Mechanics}, 744:\penalty0 404–429, 2014.
\newblock \doi{10.1017/jfm.2014.50}.

\bibitem[Perrard et~al.(2014)Perrard, Labousse, Fort, and
  Couder]{PhysRevLett.113.104101}
S.~Perrard, M.~Labousse, E.~Fort, and Y.~Couder.
\newblock Chaos driven by interfering memory.
\newblock \emph{Phys. Rev. Lett.}, 113:\penalty0 104101, Sep 2014.
\newblock \doi{10.1103/PhysRevLett.113.104101}.
\newblock URL \url{https://link.aps.org/doi/10.1103/PhysRevLett.113.104101}.

\bibitem[Labousse et~al.(2014)Labousse, Perrard, Couder, and
  Fort]{Labousse_2014}
M~Labousse, S~Perrard, Y~Couder, and E~Fort.
\newblock Build-up of macroscopic eigenstates in a memory-based constrained
  system.
\newblock \emph{New Journal of Physics}, 16\penalty0 (11):\penalty0 113027, nov
  2014.
\newblock \doi{10.1088/1367-2630/16/11/113027}.
\newblock URL \url{https://doi.org/10.1088/1367-2630/16/11/113027}.

\bibitem[Durey and Milewski(2017)]{durey_milewski_2017}
Matthew Durey and Paul~A. Milewski.
\newblock Faraday wave–droplet dynamics: discrete-time analysis.
\newblock \emph{Journal of Fluid Mechanics}, 821:\penalty0 296–329, 2017.
\newblock \doi{10.1017/jfm.2017.235}.

\bibitem[Saenz et~al.(2014)Saenz, Cristea-Platon, and Bush]{Saenz2014}
P.J. Saenz, T.~Cristea-Platon, and J.W.M. Bush.
\newblock Statistical projection effects in a hydrodynamic pilot-wave system.
\newblock \emph{Nature Phys}, 14:\penalty0 315–319, 2014.
\newblock \doi{10.1038/s41567-017-0003-x}.
\newblock URL \url{https://www.nature.com/articles/s41567-017-0003-x#citeas}.

\bibitem[Andersen et~al.(2015)Andersen, Madsen, Reichelt, Rosenlund~Ahl,
  Lautrup, Ellegaard, Levinsen, and Bohr]{Anderson15}
Anders Andersen, Jacob Madsen, Christian Reichelt, Sonja Rosenlund~Ahl, Benny
  Lautrup, Clive Ellegaard, Mogens~T. Levinsen, and Tomas Bohr.
\newblock Double-slit experiment with single wave-driven particles and its
  relation to quantum mechanics.
\newblock \emph{Phys. Rev. E}, 92:\penalty0 013006, Jul 2015.
\newblock \doi{10.1103/PhysRevE.92.013006}.
\newblock URL \url{https://link.aps.org/doi/10.1103/PhysRevE.92.013006}.

\bibitem[Pucci et~al.(2018)Pucci, Harris, Faria, and Bush]{Pucci18}
Giuseppe Pucci, Daniel Harris, Luiz Faria, and John Bush.
\newblock Walking droplets interacting with single and double slits.
\newblock \emph{Journal of Fluid Mechanics}, 835:\penalty0 1136--1156, 01 2018.
\newblock \doi{10.1017/jfm.2017.790}.

\bibitem[Frumkin et~al.(2022)Frumkin, Darrow, Bush, and Struyve]{Frumkin22}
Valeri Frumkin, David Darrow, John W.~M. Bush, and Ward Struyve.
\newblock Real surreal trajectories in pilot-wave hydrodynamics.
\newblock \emph{Phys. Rev. A}, 106:\penalty0 L010203, Jul 2022.
\newblock \doi{10.1103/PhysRevA.106.L010203}.
\newblock URL \url{https://link.aps.org/doi/10.1103/PhysRevA.106.L010203}.

\bibitem[Jamet and Drezet(2021)]{Jamet21}
Pierre Jamet and Aurélien Drezet.
\newblock A mechanical analog of bohr’s atom based on de broglie’s
  double-solution approach.
\newblock \emph{Chaos: An Interdisciplinary Journal of Nonlinear Science},
  31\penalty0 (10):\penalty0 103120, 2021.
\newblock \doi{10.1063/5.0067545}.
\newblock URL \url{https://doi.org/10.1063/5.0067545}.

\bibitem[DZYALOSHINSKII et~al.(1992)DZYALOSHINSKII, LIFSHITZ, PITAEVSKII, and
  Priestley]{DZYALOSHINSKII1992443}
I.E. DZYALOSHINSKII, E.M. LIFSHITZ, L.P. PITAEVSKII, and M.G. Priestley.
\newblock The general theory of van der waals forces.
\newblock In L.P. PITAEVSKI, editor, \emph{Perspectives in Theoretical
  Physics}, pages 443--492. Pergamon, Amsterdam, 1992.
\newblock ISBN 978-0-08-036364-6.
\newblock \doi{https://doi.org/10.1016/B978-0-08-036364-6.50039-9}.
\newblock URL
  \url{https://www.sciencedirect.com/science/article/pii/B9780080363646500399}.

\bibitem[Oza et~al.(2013)Oza, Rosales, and Bush]{oza_rosales_bush_2013}
Anand~U. Oza, Rodolfo~R. Rosales, and John W.~M. Bush.
\newblock A trajectory equation for walking droplets: hydrodynamic pilot-wave
  theory.
\newblock \emph{Journal of Fluid Mechanics}, 737:\penalty0 552–570, 2013.
\newblock \doi{10.1017/jfm.2013.581}.

\bibitem[de~Broglie(1924)]{deBroglie1924}
Louis de~Broglie.
\newblock \emph{Recherches sur la th{\'e}orie des quanta}.
\newblock PhD thesis, Migration-universit{\'e} en cours d'affectation, 1924.

\bibitem[KLEIN(1926)]{KLEIN1926}
OSKAR KLEIN.
\newblock The atomicity of electricity as a quantum theory law.
\newblock \emph{Nature}, 118\penalty0 (2971):\penalty0 516--516, Oct 1926.
\newblock ISSN 1476-4687.
\newblock \doi{10.1038/118516a0}.
\newblock URL \url{https://doi.org/10.1038/118516a0}.

\bibitem[Friedman(1922)]{Friedman1922}
A.~Friedman.
\newblock {\"U}ber die kr{\"u}mmung des raumes.
\newblock \emph{Zeitschrift f{\"u}r Physik}, 10\penalty0 (1):\penalty0
  377--386, Dec 1922.
\newblock ISSN 0044-3328.
\newblock \doi{10.1007/BF01332580}.
\newblock URL \url{https://doi.org/10.1007/BF01332580}.

\bibitem[Yu et~al.(2020)Yu, McCuller, Tse, Kijbunchoo, Barsotti, Mavalvala,
  Betzwieser, Blair, Dwyer, Effler, Evans, Fernandez-Galiana, Fritschel,
  Frolov, Matichard, McClelland, McRae, Mullavey, Sigg, Slagmolen, Whittle,
  Buikema, Chen, Corbitt, Schnabel, Abbott, Adams, Adhikari, Ananyeva, Appert,
  Arai, Areeda, Asali, Aston, Austin, Baer, Ball, Ballmer, Banagiri, Barker,
  Bartlett, Berger, Bhattacharjee, Billingsley, Biscans, Blair, Bode, Booker,
  Bork, Bramley, Brooks, Brown, Cahillane, Cannon, Chen, Ciobanu, Clara,
  Cooper, Corley, Countryman, Covas, Coyne, Datrier, Davis, Di~Fronzo, Dooley,
  Driggers, Dupej, Etzel, Evans, Feicht, Fulda, Fyffe, Giaime, Giardina,
  Godwin, Goetz, Gras, Gray, Gray, Green, Gupta, Gustafson, Gustafson, Hanks,
  Hanson, Hardwick, Hasskew, Heintze, Helmling-Cornell, Holland, Jones,
  Kandhasamy, Karki, Kasprzack, Kawabe, King, Kissel, Kumar, Landry, Lane,
  Lantz, Laxen, Lecoeuche, Leviton, Liu, Lormand, Lundgren, Macas, MacInnis,
  Macleod, Mansell, M{\'a}rka, M{\'a}rka, Martynov, Mason, Massinger, McCarthy,
  McCormick, McIver, Mendell, Merfeld, Merilh, Meylahn, Mistry, Mittleman,
  Moreno, Mow-Lowry, Mozzon, Nelson, Nguyen, Nuttall, Oberling, Oram,
  Osthelder, Ottaway, Overmier, Palamos, Parker, Payne, Pele, Perez, Pirello,
  Radkins, Ramirez, Richardson, Riles, Robertson, Rollins, Romel, Romie, Ross,
  Ryan, Sadecki, Sanchez, Sanchez, Saravanan, Savage, Schaetzl, Schofield,
  Schwartz, Sellers, Shaffer, Smith, Soni, Sorazu, Spencer, Strain, Sun,
  Szczepa{\'{n}}czyk, Thomas, Thomas, Thorne, Toland, Torrie, Traylor, Urban,
  Vajente, Valdes, Vander-Hyde, Veitch, Venkateswara, Venugopalan, Viets, Vo,
  Vorvick, Wade, Ward, Warner, Weaver, Weiss, Willke, Wipf, Xiao, Yamamoto, Yu,
  Zhang, Zucker, Zweizig, and members of~the LIGO
  Scientific~Collaboration]{Yu2020}
Haocun Yu, L.~McCuller, M.~Tse, N.~Kijbunchoo, L.~Barsotti, N.~Mavalvala,
  J.~Betzwieser, C.~D. Blair, S.~E. Dwyer, A.~Effler, M.~Evans,
  A.~Fernandez-Galiana, P.~Fritschel, V.~V. Frolov, F.~Matichard, D.~E.
  McClelland, T.~McRae, A.~Mullavey, D.~Sigg, B.~J.~J. Slagmolen, C.~Whittle,
  A.~Buikema, Y.~Chen, T.~R. Corbitt, R.~Schnabel, R.~Abbott, C.~Adams, R.~X.
  Adhikari, A.~Ananyeva, S.~Appert, K.~Arai, J.~S. Areeda, Y.~Asali, S.~M.
  Aston, C.~Austin, A.~M. Baer, M.~Ball, S.~W. Ballmer, S.~Banagiri, D.~Barker,
  J.~Bartlett, B.~K. Berger, D.~Bhattacharjee, G.~Billingsley, S.~Biscans,
  R.~M. Blair, N.~Bode, P.~Booker, R.~Bork, A.~Bramley, A.~F. Brooks, D.~D.
  Brown, C.~Cahillane, K.~C. Cannon, X.~Chen, A.~A. Ciobanu, F.~Clara, S.~J.
  Cooper, K.~R. Corley, S.~T. Countryman, P.~B. Covas, D.~C. Coyne, L.~E.~H.
  Datrier, D.~Davis, C.~Di~Fronzo, K.~L. Dooley, J.~C. Driggers, P.~Dupej,
  T.~Etzel, T.~M. Evans, J.~Feicht, P.~Fulda, M.~Fyffe, J.~A. Giaime, K.~D.
  Giardina, P.~Godwin, E.~Goetz, S.~Gras, C.~Gray, R.~Gray, A.~C. Green, Anchal
  Gupta, E.~K. Gustafson, R.~Gustafson, J.~Hanks, J.~Hanson, T.~Hardwick, R.~K.
  Hasskew, M.~C. Heintze, A.~F. Helmling-Cornell, N.~A. Holland, J.~D. Jones,
  S.~Kandhasamy, S.~Karki, M.~Kasprzack, K.~Kawabe, P.~J. King, J.~S. Kissel,
  Rahul Kumar, M.~Landry, B.~B. Lane, B.~Lantz, M.~Laxen, Y.~K. Lecoeuche,
  J.~Leviton, J.~Liu, M.~Lormand, A.~P. Lundgren, R.~Macas, M.~MacInnis, D.~M.
  Macleod, G.~L. Mansell, S.~M{\'a}rka, Z.~M{\'a}rka, D.~V. Martynov, K.~Mason,
  T.~J. Massinger, R.~McCarthy, S.~McCormick, J.~McIver, G.~Mendell,
  K.~Merfeld, E.~L. Merilh, F.~Meylahn, T.~Mistry, R.~Mittleman, G.~Moreno,
  C.~M. Mow-Lowry, S.~Mozzon, T.~J.~N. Nelson, P.~Nguyen, L.~K. Nuttall,
  J.~Oberling, Richard~J. Oram, C.~Osthelder, D.~J. Ottaway, H.~Overmier, J.~R.
  Palamos, W.~Parker, E.~Payne, A.~Pele, C.~J. Perez, M.~Pirello, H.~Radkins,
  K.~E. Ramirez, J.~W. Richardson, K.~Riles, N.~A. Robertson, J.~G. Rollins,
  C.~L. Romel, J.~H. Romie, M.~P. Ross, K.~Ryan, T.~Sadecki, E.~J. Sanchez,
  L.~E. Sanchez, T.~R. Saravanan, R.~L. Savage, D.~Schaetzl, R.~M.~S.
  Schofield, E.~Schwartz, D.~Sellers, T.~Shaffer, J.~R. Smith, S.~Soni,
  B.~Sorazu, A.~P. Spencer, K.~A. Strain, L.~Sun, M.~J. Szczepa{\'{n}}czyk,
  M.~Thomas, P.~Thomas, K.~A. Thorne, K.~Toland, C.~I. Torrie, G.~Traylor,
  A.~L. Urban, G.~Vajente, G.~Valdes, D.~C. Vander-Hyde, P.~J. Veitch,
  K.~Venkateswara, G.~Venugopalan, A.~D. Viets, T.~Vo, C.~Vorvick, M.~Wade,
  R.~L. Ward, J.~Warner, B.~Weaver, R.~Weiss, B.~Willke, C.~C. Wipf, L.~Xiao,
  H.~Yamamoto, Hang Yu, L.~Zhang, M.~E. Zucker, J.~Zweizig, and members of~the
  LIGO Scientific~Collaboration.
\newblock Quantum correlations between light and the kilogram-mass mirrors of
  ligo.
\newblock \emph{Nature}, 583\penalty0 (7814):\penalty0 43--47, Jul 2020.
\newblock ISSN 1476-4687.
\newblock \doi{10.1038/s41586-020-2420-8}.
\newblock URL \url{https://doi.org/10.1038/s41586-020-2420-8}.

\bibitem[Sen(2014)]{Debashis14}
Debashis Sen.
\newblock The uncertainty relations in quantum mechanics.
\newblock \emph{Current science}, 107:\penalty0 203 -- 218, 07 2014.
\newblock \doi{10.13140/2.1.5183.0406}.

\bibitem[Peskin(1995)]{Peskin95}
Michael~Edward Peskin.
\newblock \emph{An Introduction to Quantum Field Theory}.
\newblock Avalon Publishing, 1995.
\newblock ISBN 9780201503975.

\bibitem[Haag(1955)]{Haag:212242}
Rudolf Haag.
\newblock {On quantum field theories}.
\newblock Technical report, 1955.
\newblock URL \url{http://cds.cern.ch/record/212242}.

\bibitem[Anderson(1995)]{anderson1995computational}
J.D. Anderson.
\newblock \emph{Computational Fluid Dynamics}.
\newblock Computational Fluid Dynamics: The Basics with Applications.
  McGraw-Hill Education, 1995.
\newblock ISBN 9780070016859.
\newblock URL \url{https://books.google.com/books?id=dJceAQAAIAAJ}.

\end{thebibliography}

\end{multicols}

\end{document}